# Phase Retrieval of Vortices in Bose-Einstein Condensates

**Ron Ziv[1], Anatoly Patsyk[2], Yaakov Lumer[2],**

**Yoav Sagi[2], Yonina C. Eldar[3] and Mordechai Segev[1,2]**

[1] *Department of Electrical Engineering, Technion, Haifa 3200003, Israel*
[2] *Department of Physics and Solid State Institute, Technion, Haifa 3200003, Israel*
[3] *Department of Mathematics and Computer Science, Weizmann Institute of Science, Rehovot 7610001, Israel*
*msegev@technion.ac.il

**Abstract**

We propose and demonstrate numerically a measurement scheme for complete reconstruction of the 2D quantum wave function of a Bose-Einstein condensate, amplitude and phase, from a time-of-flight measurement. We identify a fundamental ambiguity present in the measurement of phase structures of high-symmetry excitations (e.g., vortices) and show how to overcome it by allowing for different expansion durations in different directions. We demonstrate this approach with the reconstruction of matter-wave vortices and arrays of vortices.

1. **Introduction**

Bose-Einstein condensation (BEC) is a quantum state of matter where bosonic particles form a macroscopic population in a single eigenstate. The theory [1] predicting this state waited 70 years to be explored in the lab [2, 3], a milestone achievement which launched almost three decades of fruitful research in the field of ultra-cold atoms and quantum simulators [4]. Yet, despite the progress, the commonly used measurement techniques of BECs are incomplete in the information they provide.

Imaging is at the core of measurement techniques for BECs. By shining a light through the atomic cloud and recording the shadow it casts, one can extract the density of atoms in a given state. Two imaging modes are usually available: in-situ, imaging the cloud while still inside the trap, or time-of-flight (TOF). The latter is performed by opening the trap and recording the density of atoms after expansion of the cloud [5]; it is the analog of measuring the intensity of the "far field" in optics. If the particles do not interact with one another during the expansion and the initial size of the cloud is negligible relative to the final expanded size, then the TOF image provides the momentum distribution of the cloud, which is the magnitude of the spatial Fourier transform of the wavefunction. If interactions are present but the final density is low enough such that they become negligible, the kinetic energy of the measured momentum distribution reflects the initial kinetic plus interaction energy.

These imaging modalities capture only part of the information of the state, as they measure density alone at a single point in time and in a single plane, in-situ or TOF. Yet, BECs are quantum objects, and as such they are matter waves [6], characterized by both amplitude and phase. Thus, to characterize a BEC, it is essential to obtain a complete map of their amplitude and phase everywhere in space, as they evolve. Accordingly, relying on these two modalities, innovative methods were developed to characterize these states, such as Bragg spectroscopy for momentum measurements [7,8], real-time probing [9,10], Josephson effects and Raman superradiance [11,12]. For measuring the phase of the condensate wavefunction, interference can be exploited – either between different parts of the same BEC [13] or between different BECs [14]. However, interference measurements of cold atoms present significant experimental challenges [15], as they require coherently splitting and recombining a BEC. In addition, such measurements are destructive; hence, each measurement captures a single realization of the experiment. This raises a natural question: **Can the phase and amplitude of a BEC wavefunction be recovered without atomic interference of two clouds** [16]? This question is related to the well-known phase retrieval problem from optics but with some important differences which we highlight below.

Traditionally, the phase retrieval problem is defined as the recovery of an object, amplitude and phase, from the magnitude of its Fourier transform [17]. This problem arises naturally in coherent optical imaging as the far-field of an object (or the field at the focal plane of a lens) is proportional to the Fourier transform of the object that one wishes to image. As most detectors, such as digital cameras, only measure the field's intensity, the information measured is the magnitude (squared) of the Fourier transform, losing the phase information. Since the phase structure of a coherent light beam is also embedded in the far-field diffraction pattern, one can try and recover it computationally from the intensity measurement in the far-field. Under suitable conditions and up to certain ambiguities, one can often recover a suitable phase function in an iterative fashion by utilizing prior knowledge about the state to be reconstructed (e.g., a finite "support" - the region within which the image is contained). A well-known example is Fienup's algorithm [18] from the late seventies, which is commonly used in optics for phase retrieval. Moreover, under sufficient conditions on the image, a unique solution can sometimes be guaranteed [19]. Additional prior knowledge, such as sparsity, can result in improvements in terms of the number of sample points (resolution of the detector), noise robustness and convergence rate of the algorithm, even of 1D objects, as was explored in many works [20–24]. In recent years, the phase retrieval problem has been studied extensively, and several measurement schemes and recovery algorithms have been proposed and demonstrated, facilitating recovery of the phase from various forms of generalized Fourier measurements [25–28].

The setting of phase-retrieval in optical imaging is very similar to TOF in BEC measurements, as the information of the initial state is embedded in some far-field plane. By the same logic, if the initial quantum state of the BEC includes a phase structure, it affects the evolution to the plane where TOF measurements are carried out; hence one can try and extract it from the TOF measurements. However, unlike the traditional phase retrieval problem from optics, where the relation between the electromagnetic field in the image plane and the field in the far-field is a simple Fourier transform, in the expansion dynamics of a BEC the relation is more complicated. Specifically, the particles constituting the BEC interact with one another; hence the propagation from the BEC to the measurement plane is no longer ballistic: it does not follow a simple linear relation (such as a Fourier transform) but instead is governed by a nonlinear evolution equation. Also unlike in optics, as noted earlier, BEC measurements are typically destructive, and if non-destructive techniques [5] are employed they come at the expense of the signal to noise ratio. Additionally, BECs have a limited time within which they still act as a coherent entity when the trap is removed, which sets a limit on the path and phase difference in interference measurements [29].

The concept of phase retrieval has been proposed for atomic BEC TOF measurements before [30,31], and even for light emission from a plasmonic BEC [32]. However, these methods do not handle well complex phase structures such as vortices, which naturally create ambiguities in a phase retrieval process based on density TOF measurements, due to symmetries in propagation. Pointedly, these ambiguities cannot be resolved by a phase-retrieval algorithm. For example, employing the Fienup phase-retrieval methodology or a more modern technique [17] cannot reveal the rotation direction of a vortex such as in Ref. [33], or an array of vortices, as has often been studied in BEC experiments [34,35]. The issue of unraveling the directionality and phase of vortices in a BEC is extremely important for multiple reasons, for example - to determine the vortex quantum number (i.e., its topological charge). Being able to measure the topological charge of all the vortices is crucial in the study of many physical phenomena, such as the Berezinskii-Kosterlitz-Thouless phase-transition [35], where pairs of counter-rotating vortices appear spontaneously. Understanding the superfluid dynamics of the BEC [10] also requires mapping the quantum numbers of vortices. In addition, vortices can be used to carry quantum information [36], which may be encoded by the topological charge. In a similar vein, vortices play a major role in the formation of matter-wave vortex-solitons [37–39], where the spatial dependence of the phase is a key feature [40,41]. Therefore, the recovery of vortex structures in a BEC is crucial and has many far-reaching implications for basic and applied science.

Over the years, various experimental methods for studying vortices in atomic gases were developed. For instance, Seo et al. [42] employed Bragg spectroscopy with counter propagating beams for sign-detection of vortices from TOF images. Likewise, Haljan et al. [43] were the first to distinguish between a vortex and anti-vortex by tilting the condensate and inducing spin precession. Other closely related methods followed [44,45]. Serafini et al. [46] studied dynamics of single vortex and two vortices in a 3D BEC by periodically outcoupling and imaging a fraction of the condensate atoms, and the authors in Refs. [39,47] studied the case of a single solitonic vortex in an elongated BEC, by triaxial TOF density imaging, relating the twisting to the vortex sign while also employing Bragg interferometry for detection of vortex fork dislocations. However, all of these novel methods cannot fully recover the spatial phase structure of the condensate's wavefunction. Therefore, extending phase retrieval methods to BECs, specifically in the case of vortex structures, can bring the benefits of algorithmic approaches, characterizing the density and phase of a condensate with a single simple measurement, into the realm of BECs.

Here, we propose and demonstrate in numerical simulations a scheme for the complete characterization of the 2D quantum wavefunction imprinted on a Bose-Einstein condensate from TOF measurements alone, including wavefunctions containing vortices. Our proposed measurement scheme is based on a simple variation to TOF measurements and does not require interference with another condensate. The variation breaks the radial symmetry in propagation, which is the source of ambiguities in the measurement of vortices, and otherwise cannot be lifted by any algorithm based on TOF imaging. Our measurement scheme resolves these ambiguities and facilitates the recovery of single vortices and of vortex arrays, including their directionality, without atomic interference or multiple measurement planes. Moreover, the scheme is general and does not assume any prior information on the wavefunction, trap or any other physical parameter, it only assumes TOF propagation based on the Gross-Pitaevskii equation (GPE). Accordingly, since the evolution of BECs is nonlinear, described by the GPE, our algorithm is based on nonlinear dynamical evolution rather than on the simple Fourier transform used in linear phase-retrieval problems.

## 2. Background

### *2.1 Mean field description*

The object we wish to reconstruct is the 2D wave function imprinted on a Bose-Einstein Condensate. The dynamics of the state of N particles are given by the many-body Hamiltonian. Under appropriate conditions, a mean-field approximation is valid, and the many-body Hamiltonian describing the ensemble of particles is reduced to the Gross-Pitaevskii equation [48], which is functionally equal to the non-linear Schrödinger-type equation:

$$\left(-\frac{\hbar^2}{2m}\nabla^2 + V(\boldsymbol{r}) + U_0|\psi(\boldsymbol{r},t)|^2\right)\psi(\boldsymbol{r},t) = i\hbar\partial_t\psi(\boldsymbol{r},t) \qquad 1$$

where $V(\mathbf{r})$ is the confining potential of the trap as a function of coordinate $\mathbf{r}$, $V$ typically being a real function, and $U_0 > 0$ is the nonlinear coefficient that represents repulsion between particles due to interactions. Usually, it is reasonable to assume that the wavefunction is initially localized by the confining potential; hence knowledge of the region within which the wavefunction density is nonzero (the "support" of the function) is tight.

An alternative mathematical description of the state above highlights the relation between the phase of the wavefunction and the flow of quantum gas. Through a hydrodynamics description [49], one can show that the following holds:

$$\psi(\mathbf{r},\theta) = f(\mathbf{r},\theta)e^{i\phi(\mathbf{r},\theta)}$$
$$\mathbf{v} = \frac{\hbar}{m}\nabla\phi \qquad 2$$

where $\psi(\mathbf{r},\theta)$ is the wavefunction describing the condensate in terms of its amplitude and phase and $\mathbf{v}$ is the expectation value of the velocity operator using the condensate wavefunction. One special type of flow is that of a vortex, that is, a field that circulates around a point. Because the phase (modulus $2\pi$) must be continuous, a vortex flow corresponds to a phase function of the functional form $\phi(\mathbf{r},\theta) = n\theta$, where $n$ is an integer. As a result, the rotation speed of vortices is always quantized in a quantum gas. Due to this quantization, one can assign a topological quantity to the vortex, commonly referred to as the topological charge ($n$, in our case). In a similar fashion, optical vortex beams are electromagnetic waves that carry orbital angular momentum (OAM), where $n$ is the order of the OAM and its sign gives its direction.

### *2.2 Time of Flight measurement*

A time-of-flight measurement is performed by opening the trap confining the condensate and allowing the atomic cloud to expand freely. After some waiting time, collimated light is launched through the cloud. The light is partially absorbed by the atoms and re-scattered to all directions, which cast a shadow of the atomic cloud onto the beam front. At light intensities below the saturation intensity, the absorption coefficient is linearly proportional to the density of atoms. Mathematically, this procedure translates to measuring $|\psi(\mathbf{r},T)|^2$ at some time $T$ after the trap is removed and the atomic cloud is allowed to evolve freely, i.e., with no potential present, setting $V(\mathbf{r})$ to zero in Eq. 1.

An important observation is that, when the trap is turned off, the potential-free Hamiltonian conserves orbital angular momentum and hence also conserves the total OAM of the state, even for the nonlinear evolution of the GPE. In the context of BECs containing vortices, this implies that the condensate conserves OAM in the free propagation of the TOF measurement. As the phase of a vortex is singular at its core, it results in a zero density there.

Due to the symmetries in Eq. 1, this zero density core persists throughout the propagation and hence is also present in the TOF measurement. This implies that the presence of a vortex can be detected by locating the zero density point. Alas, as the OAM is manifested in the phase of the state, it means that a state with opposite vortex direction results in identical density measurement; this means one cannot distinguish the directionality of the vortex flow from this measurement, as can be seen in Fig 1. Likewise, in a BEC with multiple vortices, locating the zero-density points does not reveal the rotation directions of the individual vortices. This fact presents a problem for the phase retrieval algorithm. For any reconstruction algorithm, when one measurement corresponds to two (or more) different inputs - the reconstruction problem is ill-posed and contains an ambiguity preventing recovery of the original input. While some ambiguities are trivial and not important, such as global phase, this type of ambiguity has a physical meaning, and hence lifting it is crucial for successful operation.

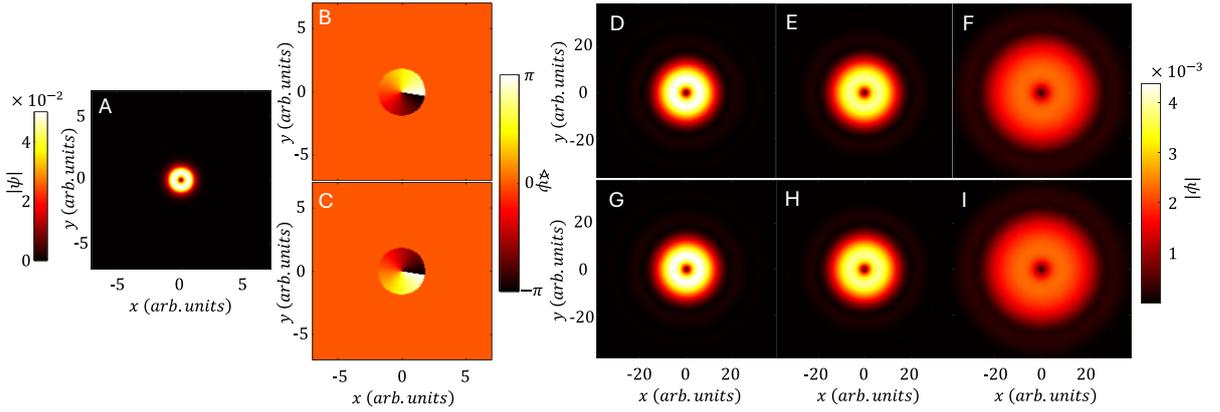

*Fig 1.* (A) Amplitude distribution of single $m=\pm 1$ vortex state in arbitrary unit. (B) phase structure of vortex state $m=1$. (C) phase structure of vortex state $m=-1$. (D-I) Top row: TOF measurement of state (B) with linear(D), "weak"(E) and "strong" nonlinear(F) propogation. Bottom row: TOF measurement for state (C) with linear (G), "weak"(H) and "strong" nonlinear (I) propogation. See Appendix B for simulation and propagation regimes details.

### *2.3 Phase Retrieval*

Phase retrieval is the mathematical problem of reconstructing a function from the magnitude of its Fourier transform. Without the phase of the Fourier components - information is lost [50], and the transform is not bijective. Hence, recovering the initial object amounts to retrieving the phase in the Fourier plane. The phase retrieval problem has been studied extensivity in the past and in recent years as well, bringing forth theoretical results guaranteeing uniqueness and stability under various constraints (such as prior information) on the input and new classes of measurements [19] along with new algorithms for reconstruction [21,22].

The most common phase retrieval algorithms are iterative [51], and are derived either by solving an underlying optimization problem, or by using alternating projections based on the following working principle: As we have two relevant planes, the object plane and the Fourier plane, one can impose the information one has at each plane and iterate between the two planes. For instance, one starts by drawing a random initial wavefunction guess and propagate it by Fourier transform to the far-field. Here, the magnitude is replaced with the measured Fourier magnitude. The field is then propagated back (inverse Fourier transform) to the object plane, where constraints are imposed, such as support, sparsity and more.

Traditional phase retrieval methods, as noted above, were developed and used based on Fourier propagation. However, TOF measurements of BECs follow Eq. 1, which is nonlinear, and the measurements are not of the Fourier-transform squared. Rather, the measurements are taken over the atom density $|\psi(\mathbf{r},t)|^2$ sometime $T$ after the trap is removed. Nevertheless, the underlying principle of these algorithms can still be applied to different mechanisms of evolution, linear or even nonlinear, as we show below. This idea has been proposed in nonlinear optics [52] and also for BECs [30], but has thus far never been demonstrated experimentally with cold atoms. However, while these methods can work well, for measurements with ambiguities associated with the inherent symmetries of the evolution according to the GPE, Eq. 1 - they fail. In what follows, we describe our methodology of phase-retrieval of BECs, focusing on changing the measurement scheme for the challenging task of phase-retrieval of wavefunctions containing vortices.

## 3 Method

### *3.1 Augmented TOF*

The reason an ambiguity arises in the measurement of the TOF images from vortices is radial symmetry of the GPE, which results in the same amplitude for a right-handed vortex and a left-handed one. In order to alleviate this ambiguity, we propose to break this symmetry in the TOF propagation. We achieve this by a simple and feasible adaption to the TOF measurement. Instead of opening the BEC trap in the x-y axes simultaneously, we open the trap in succession: one axis first and after sufficient evolution, the second axis. That is, there are two times relevant for propagation, $T_1$ the propagation time under a partial potential active in one axis only, followed by free propagation for a time $T_2$, and then performing the measurement which yields the density of atoms. This procedure removes the radial symmetry between positive and negative singly-charged vortices. However, as we see in our simulations described in Section 3.4, this procedure can remove the ambiguities also for intricate states containing many vortices. In this TOF, $T_2$ should be long enough such that the dynamics create enough mixing of phase information within this time interval. This is in analogy to the regular Fourier propagation, where one wants sufficient propagation to approximate the Fourier transform. Henceforth we refer to this methodology as augmented TOF. Our augmented TOF phase-retrieval methodology is conceptually similar to breaking the propagation symmetry in linear optics by an intermediate cylindrical lens [53].

Figure 2 shows the same states from Fig 1 after augmented TOF measurements, instead of the standard TOF measurements shown in Fig. 1. As shown in Fig. 2, unlike the regular TOF measurement that yields the same result for vortex of order 1 and -1 (Fig. 1), the augmented TOF produces different patterns regardless of the interaction strength.

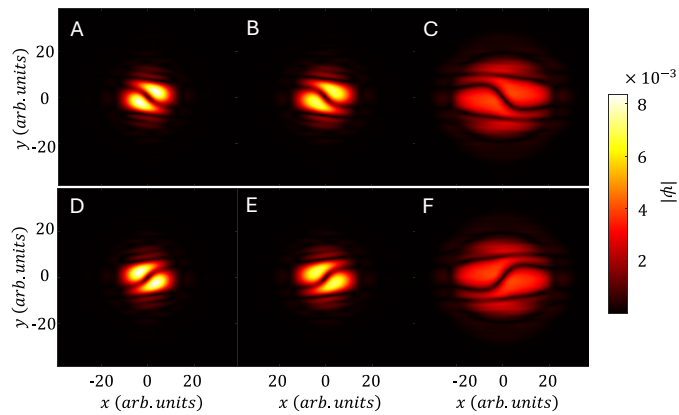

**Fig 2.** Top row: Augmented TOF measurement for vortex states of $m = 1$ with linear(A), weak(B) and strong nonlinear(C) propogation. Bottom row: Augmented TOF measurement for vortex states of $m = -1$ with the same propogation parameters for "linear"(D), "weak"(E) and "strong" nonlinear(F) propogation.

### *3.2 GPE-Phase Retrieval Algorithm*

We first address the phase retrieval problem for the nonlinear propagation represented by the GPE. The problem is similar to the common phase retrieval. Thus, the algorithm for this nonlinear evolution is almost identical to common phase retrieval algorithms, with appropriate modification. As the propagation is no longer linear, we modify existing algorithms by replacing the Fourier propagation with GPE-based propagation, where we include in this propagation the different times under the partial trapping and free expansion. This is done by numerically solving the GPE for each iteration backward and forward. In principle, for methods requiring gradient computations, we would need to adapt the gradient as well. Instead, we focus on Fienup-type methods which rely on projections between planes so that we only modify the forward model while the rest of the algorithm steps remain unchanged.

Our numerical solver is the split-step method or Beam Propagation Model [54]. As in-situ atom-density measurements are available for the BEC (sometimes at low resolution), we impose a magnitude constraint instead of support constraint in the in-situ plane. This scheme is sketched in Fig 3, where $\psi_i$ is the wavefunction estimated at the trap plane in iteration $i$ and $\widetilde{\psi_i}$ is the wavefunction estimate at the measurement plane, while wavefunctions without iteration index correspond to the measured images at each plane. Reconstruction results for this algorithm are presented in section 3.4. The algorithm in Fig 3 is the same for the case of the BEC with our proposed augmented TOF measurement. The only difference is that the GPE propagation (and its inverse) is done in two steps, numerical propagation under the partial potential for a time $T_1$ followed by numerical free propagation with no potential for a time $T_2$. The measurement constraint is changed into the augmented TOF instead of TOF.

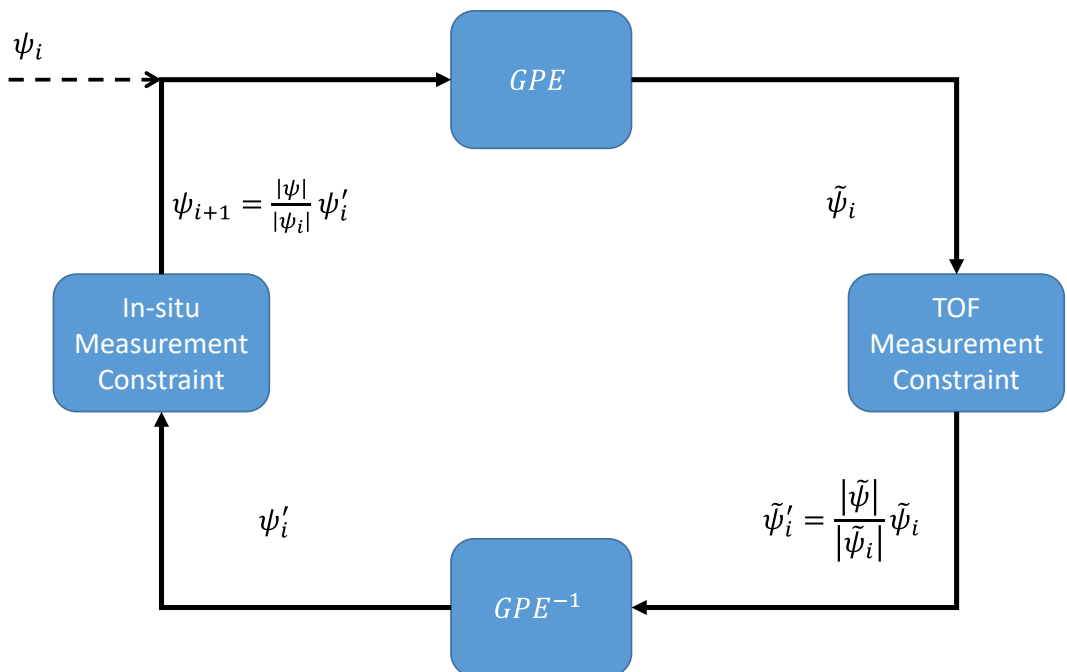

**Fig 3.** Graphical depiction of the iterative phase retrieval algorithm for the augmented TOF measurements of BEC.

### 3.3 $T_1$ *Propagation Time*

As the augmented TOF scheme introduces an extra parameter, the time under the partial trap, an immediate question arises: what is the optimal propagation time under the partial trap? to select a possible $T_1$, we choose to optimize the maximal norm difference between a vortex state of order 1 and order -1. That is:

$$\max_{T_1} \||\psi_1(x,y,T_1+T_2)|^2 - |\psi_{-1}(x,y,T_1+T_2)|^2\| \qquad 3$$

Where $\psi_{\pm 1}(x,y,T_1+T_2)$ is a vortex state with an order of $\pm 1$, as a function of propagation time under the partial trap $T_1$, followed by free propagation of time $T_2$. The intuition behind this selection is maximizing the difference between the measured images in the augmented TOF for the ambiguous condensates.

We relax the above problem by solving for the linear regime of the GPE and taking $T_2$ such that a far-field approximation is valid. Under these conditions, we show (Supp A.) that the norm difference is approximately proportional to:

$$\||\psi_1(x,y,T_1+T_2)|^2 - |\psi_{-1}(x,y,T_1+T_2)|^2\| \propto \frac{\left|\sin\left(\frac{\Delta E T_1}{\hbar}\right)\right|}{\sqrt{T_2(T_1+T_2)}} \qquad 4$$

where $\Delta E$ is the energy difference between the first and second modes of the trap. Note that this result is independent of the norm choice and as such the notation was left general. By setting $\frac{\Delta E T_1}{\hbar} = \frac{\pi}{2}$ we can expect a maximal difference between the two states. We validate this result numerically, as shown in Fig. 4. We note that this result can be computed directly from the parameters of the system, with no additional measurement required for the calculation.

In general, these results can be further developed to deal with high-order vortices and more intricate phase structures, such as lattice of vortices, and to optimize over the reconstruction error of the algorithm rather than the measurement difference. We leave these improvements for future work.

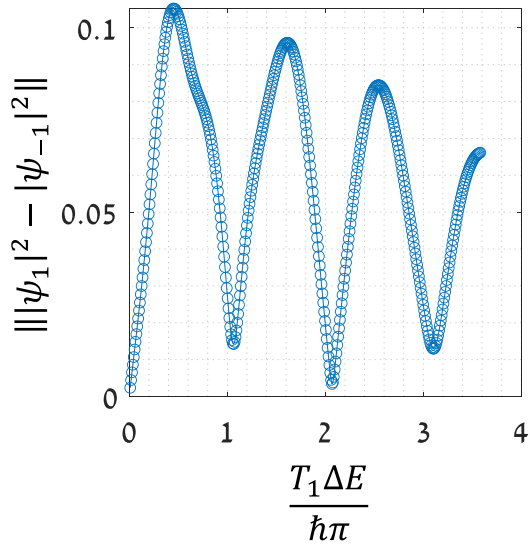

**Fig 4.** Norm difference between augmented TOF measurements of vortex states $m=1$ and $m=-1$ as a function of propagation time $T_1$. Time axis is scaled by the energy difference between the second and first modes of the trap.

### 3.4 Reconstruction

To test the proposed method, we simulate a more complex class of wavefunctions and show their respective reconstructions. As an example, we consider a 3x3 lattice of order-one vortices. Each vortex having a random order sign and relative phase to the others. Each vortex by itself is a stationary order-one vortex state of the GPE equation with a single site potential well. We test the performance of our method on an ensemble of 150 such lattices. We measure the augmented time of flight and in-situ measurements for 150 wavefunctions and different interaction strengths and reconstruct the wavefunction based on the algorithm described in Section 3.2. Unlike the case of the single vortex state, where we analytically show there is ambiguity in the TOF measurement, for the lattice case this is not that simple because only the total topological charge is conserved. For example, single-charge vortices can merge and/or multi-charged vortices can split during propagation, hence local charge is not conserved. As expected, our numerical study reveals ambiguity also in attempting to reconstruct the vortex lattice based on TOF measurements. This can be seen in Fig 5, where we attempt to reconstruct an initial wavefunction based on TOF measurement and on augmented TOF measurement. As shown there, for the TOF case, the algorithm converges into an erroneous phase signal, yet as shown in the figure, the augmented TOF yields the correct reconstructed wavefunction. These numerical experiments lead us to conjecture that some vortex lattices contain ambiguities in TOF measurements of BECs, which can be overcome by the methodology of augmented TOF measurements and the accompanying algorithm.

To test the reconstruction further, we simulate measurements in the presence of white Gaussian noise of a varying degree of signal to noise ratio (SNR) in the range $[60dB - 0dB]$, for an ensemble of 10 random vortex lattices, similar our previous test. While more tests are required to validate noise robustness, we numerically observe that we are able to reconstruct vortex lattices even in the presence of relatively high noise, as can be seen in Fig 6. Note that, while some noise is still present in the final phase reconstruction, it is centered in regions of low magnitude (i.e., low signal strength). Importantly, the direction of each vortex and its relative phase with respect to others can be easily distinguished. Remarkably, the final reconstructed augmented TOF measurement shows considerable noise reduction, thereby indicating good performance.

Last, to illustrate that the augmented TOF is not limited to vortex lattices and can recover complete spatial phase information, as is the case with traditional phase retrieval, we consider the case of random phase imprinted on a ground state of BEC inside a harmonic trap. We recover an ensemble of 10 random phase patterns. An example of a specific reconstruction is shown in Fig 7 showcasing virtually perfect recovery of the random phase based on the augmented TOF measurements.

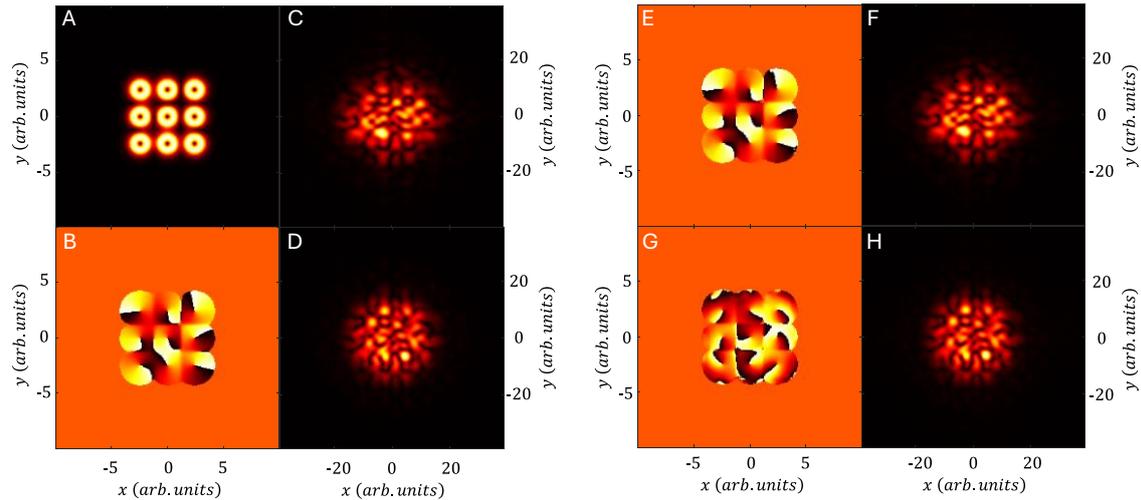

**Fig 5.** (A) Amplitude of wavefunction for reconstruction and (B) its phase, (C) is the augmented TOF and (D) is the original TOF. In (E), we present the phase reconstructed with the augmented TOF scheme and its corresponding phase retrieval algorithm. The far field result is shown in (F). In (G), an erroneous phase reconstruction is shown using the original TOF scheme, producing the "measured" far field measurement (H) but with a wrong object phase.

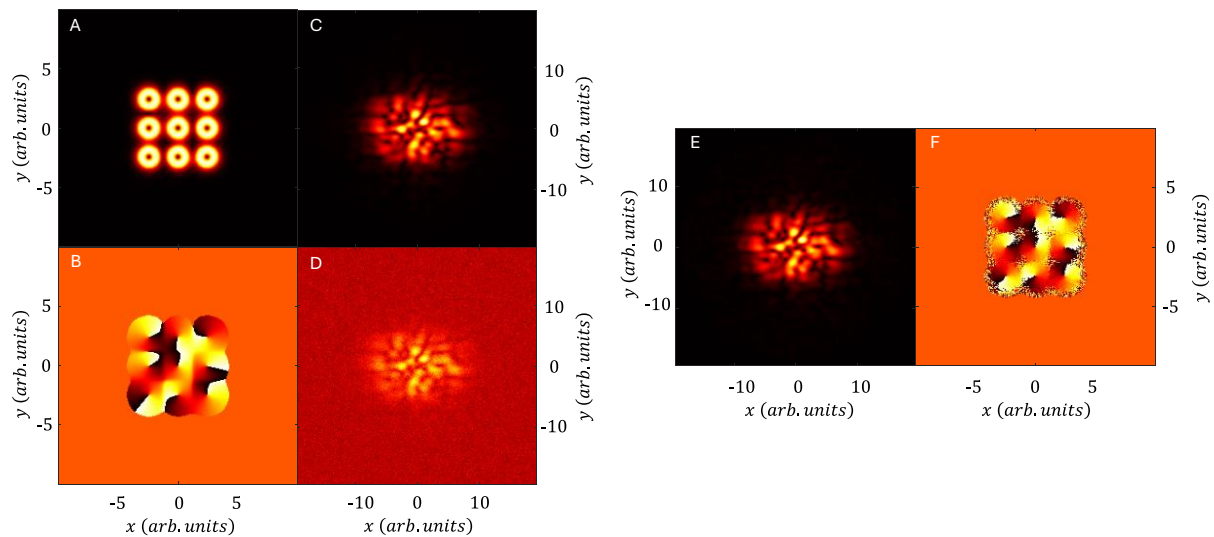

**Fig 6.** (A,B) Amplitude and phase of the reconstructed wavefunction, respectively. (C,D) Calculated atom density corresponding to augmented TOF measurement of (A,B), without appreciable noise and with added gaussian white noise, respectively. The noise in (D) is very strong, corresponding to SNR 0dB, yet the measurement scheme and the algorithm facilitate correct reconstruction, amplitude and phase, as shown in (E) and (F), respectively.

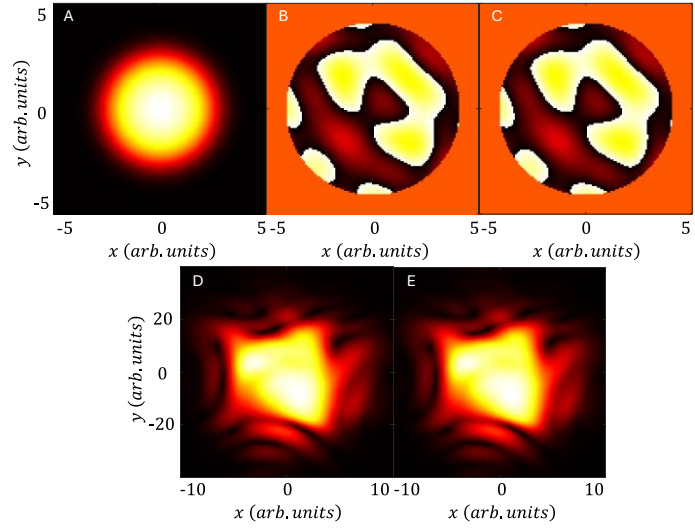

**Fig 7.** (A,B) Amplitude and random phase pattern of the wavefunction, respectively. (C) reconstructed phase based on (D) – which is the calculated augmented TOF measurement of (A,B). (E) Augmented TOF of reconstructed wavefunction. Note the asymmetric aspect ratio in the augmented TOF images.

## 4 Discussion and Outlook

Our results show that the problem of reconstructing the 2D phase structure of a BEC from TOF measurements is analogous to the phase retrieval problem in optics, with the distinction that the evolution of a BEC is nonlinear. We have shown that vortices and periodic arrays of vortices pose a problem that phase-retrieval (linear and nonlinear) from TOF (or far-field) measurements cannot resolve due to ambiguities. We proposed a method to resolve this ambiguity problem through a simple augmentation of the measurement protocol of a BEC by a two-stage procedure: the trap is opened first in one direction and only after some time in the other direction. This procedure was shown to empirically remove the ambiguities and, together with the proper phase-retrieval algorithm for nonlinear evolution, facilitates recovery of complex quantum wavefunctions containing vortices and lattices of vortices. As vortices are a ubiquitous phenomenon with important implications for quantum technology and basic science, their reconstruction – including their helicity (topological charge) is crucial.

Experimentally, having distinct control over the two directions necessitates building the trap using two separate components, each influencing one direction. One method to achieve this involves intersecting two far-off-resonance laser beams propagating in orthogonal directions and switching them off at different times during the modified TOF measurement. Another experimental aspect, relating to the noisy reconstructions presented, is the imaging resolution and its effects on the reconstruction capabilities. Intuitively, the resolution should be such that all relevant density features are captured in the measurement. As the cloud expands, this creates a trade-off between imaging resolution, propagation time and noise level. While not pursued in this work, a study of the interplay between the above is an interesting and important direction of research.

As the algorithm outlined uses in-situ and augmented TOF density images, this might present a challenge in cases where nondestructive imaging is not available, as these measurements cannot be obtained in a single shot. In practice, this can be resolved by several approaches. If the initial state production is highly reliable, one can combine measurements from different shots. Otherwise, like with other phase retrieval algorithmic variants, an alternative can be to exchange the in-situ magnitude constraint, which requires a measurement, with a support constraint defined by the trap support.

We analytically found the optimal propagation time under the partial trap for maximizing the measurement difference in the case of linear propagation and a single vortex and related it to the energy difference between the modes of the trap. In the context of measurement design, one possible future direction can be potential engineering. That is, instead of turning off the trap in stages in separate axes, engineering a time-dependent potential that will optimize signal recovery for specific classes of wavefunctions.

On the algorithmic side, the proof of concept we proposed utilized a basic iterative Fienup-type algorithm and shows there is sufficient information in the measurement for reconstruction in the cases we examined. While not pursued in this work, more advanced and novel algorithms can be adapted to the case of the BEC (especially the nonlinear propagation) to allow more robust recovery in the presence of noise and analysis of the conditions for uniqueness and correctness.

We would like to emphasize that the only assumption of our method is that the propagation of the state follows the GPE, i.e., the mean-field approximation, in 2D. This means that the cloud is effectively in 2D configuration in space and evolution under the full trap does not strictly need to follow the GPE model; only the propagation of the measurement must follow the GPE. This point also hints at future directions of incorporating more complex and advanced models into the retrieval problem, such as the Bogoliubov approximation [55] and other model of BEC in the presence of a thermal cloud [56]. Another potential avenue is exploring the effects of 3D cloud configurations on the 2D recovery and regimes of validity outside the assumption of effective 2D dynamics. This also hints at the feasibility of 3D reconstruction, similar to tomographic reconstruction.

Last but not least, we note that thus far, algorithmic phase-retrieval of the quantum wavefunctions of cold atoms has not been demonstrated in experiments. We are now pursuing this concept with an experimental group [57]. Clearly, succeeding in recovering the quantum wavefunction from TOF measurements will introduce a new powerful tool for experiments with cold atoms and will also apply to cold molecules [58], removing the excessively hard requirement for atom interference for unraveling the phase.

## References

[1] A. Einstein, Zur quantentheorie des idealen gases, in Albert Einstein: Akademie-Vorträge (John Wiley & Sons, Ltd, Hoboken, NJ, 2005), pp. 258–266.
[2] M. H. Anderson, J. R. Ensher, M. R. Matthews, C. E. Wieman, and E. A. Cornell, *Observation of Bose-Einstein Condensation in a Dilute Atomic Vapor*, Science **269**, 198 (1995).
[3] K. B. Davis, M.-O. Mewes, M. R. Andrews, N. J. van Druten, D. S. Durfee, D. M. Kurn, and W. Ketterle, *Bose-Einstein Condensation in a Gas of Sodium Atoms*, Physical Review Letters **75**, 3969 (1995).
[4] D. W. Snoke, N. P. Proukakis, T. Giamarchi, and P. B. Littlewood, Universality and Bose–Einstein condensation: Perspectives on recent work, in Universal Themes of Bose–Einstein Condensation (Cambridge University Press, Cambridge, England, 2017).
[5] W. Ketterle, D. S. Durfee, and D. M. Stamper-Kurn, Making, probing and understanding Bose-Einstein condensates, in atomic gases, in Proceedings of the International School of Physics "Enrico Fermi" (IOS Press, Amsterdam, Netherlands, 1999), pp. 67–176.

[6]  M. R. Andrews, C. G. Townsend, H.-J. Miesner, D. S. Durfee, D. M. Kurn, and W. Ketterle, *Observation of Interference between Two Bose Condensates*, Science **275**, 637 (1997).

[7]  J. Stenger, S. Inouye, A. P. Chikkatur, D. M. Stamper-Kurn, D. E. Pritchard, and W. Ketterle, *Bragg Spectroscopy of a Bose-Einstein Condensate*, Phys. Rev. Lett. **82**, 4569 (1999).

[8]  S. Richard, F. Gerbier, J. H. Thywissen, M. Hugbart, P. Bouyer, and A. Aspect, *Momentum Spectroscopy of 1D Phase Fluctuations in Bose-Einstein Condensates*, Phys. Rev. Lett. **91**, 010405 (2003).

[9]  A. Ramanathan, S. R. Muniz, K. C. Wright, R. P. Anderson, W. D. Phillips, K. Helmerson, and G. K. Campbell, *Partial-Transfer Absorption Imaging: A Versatile Technique for Optimal Imaging of Ultracold Gases*, Review of Scientific Instruments **83**, 083119 (2012).

[10] D. V. Freilich, D. M. Bianchi, A. M. Kaufman, T. K. Langin, and D. S. Hall, *Real-Time Dynamics of Single Vortex Lines and Vortex Dipoles in a Bose-Einstein Condensate*, Science **329**, 1182 (2010).

[11] S. Levy, E. Lahoud, I. Shomroni, and J. Steinhauer, *The a.c. and d.c. Josephson Effects in a Bose–Einstein Condensate*, Nature **449**, 7162 (2007).

[12] L. E. Sadler, J. M. Higbie, S. R. Leslie, M. Vengalattore, and D. M. Stamper-Kurn, *Coherence-Enhanced Imaging of a Degenerate Bose-Einstein Gas*, Phys. Rev. Lett. **98**, 110401 (2007).

[13] E. W. Hagley, L. Deng, M. Kozuma, M. Trippenbach, Y. B. Band, M. Edwards, P. S. Julienne, K. Helmerson, S. L. Rolston, and W. D. Phillips, *Measurement of the Coherence of a Bose-Einstein Condensate*, Physical Review Letters **83**, 3112 (1999).

[14] M. E. Zawadzki, P. F. Griffin, E. Riis, and A. S. Arnold, *Spatial Interference from Well-Separated Split Condensates*, Physical Review A **81**, 043608 (2010).

[15] Y. Torii, Y. Suzuki, M. Kozuma, T. Sugiura, T. Kuga, L. Deng, and E. W. Hagley, *Mach-Zehnder Bragg Interferometer for a Bose-Einstein Condensate*, Phys. Rev. A **61**, 041602 (2000).

[16] R. Ziv, Y. Sagi, Y. C. Eldar, and M. Segev, Phase retrieval of vortices in Bose-Einstein condensates, in 2021 Conference on Lasers and Electro-Optics (CLEO) (IEEE, San Jose, CA, 021), pp. 1–2.

[17] Y. Shechtman, Y. C. Eldar, O. Cohen, H. N. Chapman, J. Miao, and M. Segev, *Phase Retrieval with Application to Optical Imaging: A Contemporary Overview*, IEEE Signal Processing Magazine **32**, 87 (2015).

[18] J. R. Fienup, *Reconstruction of an Object from the Modulus of Its Fourier Transform*, Optics Letters **3**, 27 (1978).

[19] T. Bendory, R. Beinert, and Y. C. Eldar, Fourier phase retrieval: Uniqueness and algorithms, in Compressed Sensing and Its Applications (Springer, Cham, Switzerland, 2017), pp. 55–91.

[20] Y. Shechtman, Y. C. Eldar, A. Szameit, and M. Segev, Sparsity based sub-wavelength imaging with partially incoherent light via quadratic compressed sensing, Opt. Exp. **19**, 14807 (2011).

[21] Y. Shechtman, A. Beck, and Y. C. Eldar, *GESPAR: Efficient Phase Retrieval of Sparse Signals*, IEEE Transactions on Signal Processing **62**, 928 (2014).

[22] G. Wang, G. B. Giannakis, and Y. C. Eldar, *Solving Systems of Random Quadratic Equations via Truncated Amplitude Flow*, IEEE Transactions on Information Theory **64**, 773 (2017).

[23] G. Wang, L. Zhang, G. B. Giannakis, M. Akçakaya, and J. Chen, *Sparse Phase Retrieval via Truncated Amplitude Flow*, IEEE Transactions on Signal Processing **66**, 479 (2017).

[24] P. Schniter and S. Rangan, *Compressive Phase Retrieval via Generalized Approximate Message Passing*, IEEE Transactions on Signal Processing **63**, 1043 (2014).

[25] T. Bendory, D. Edidin, and Y. C. Eldar, *On Signal Reconstruction from FROG Measurements*, Applied and Computational Harmonic Analysis **48**, 1030 (2020).

[26] T. Bendory, Y. C. Eldar, and N. Boumal, *Non-Convex Phase Retrieval from STFT Measurements*, IEEE Transactions on Information Theory **64**, 467 (2017).

[27] K. Huang, Y. C. Eldar, and N. D. Sidiropoulos, *Phase Retrieval from 1D Fourier Measurements: Convexity, Uniqueness, and Algorithms*, IEEE Transactions on Signal Processing **64**, 6105 (2016).

[28] K. Jaganathan, Y. C. Eldar, and B. Hassibi, *STFT Phase Retrieval: Uniqueness Guarantees and Recovery Algorithms*, IEEE Journal of Selected Topics in Signal Processing **10**, 770 (2016).

[29] G.-B. Jo, J.-H. Choi, C. A. Christensen, Y.-R. Lee, T. A. Pasquini, W. Ketterle, and D. E. Pritchard, *Matter-Wave Interferometry with Phase Fluctuating Bose-Einstein Condensates*, Phys. Rev. Lett. **99**, 240406 (2007).
[30] A. Kosior and K. Sacha, *Condensate Phase Microscopy*, Phys. Rev. Lett. **112**, 045302 (2014).
[31] D. Meiser and P. Meystre, *Reconstruction of the Phase of Matter-Wave Fields Using a Momentum-Resolved Cross-Correlation Technique*, Phys. Rev. A **72**, 023605 (2005).
[32] J. M. Taskinen, P. Kliuiev, A. J. Moilanen, and P. Törmä, *Polarization and Phase Textures in Lattice Plasmon Condensates*, Nano Lett. **21**, 5262 (2021).
[33] M. R. Matthews, B. P. Anderson, P. C. Haljan, D. S. Hall, C. E. Wieman, and E. A. Cornell, Vortices in a Bose-Einstein condensate, Phys. Rev. Lett. **83**, 2498 (1999).
[34] J. R. Abo-Shaeer, C. Raman, J. M. Vogels, and W. Ketterle, Observation of vortex lattices in Bose-Einstein condensates, Science **292**, 476 (2001).
[35] C. Lobo, A. Sinatra, and Y. Castin, *Vortex Lattice Formation in Bose-Einstein Condensates*, Physical Review Letters **92**, 020403 (2004).
[36] K. T. Kapale and J. P. Dowling, *Vortex Phase Qubit: Generating Arbitrary, Counterrotating, Coherent Superpositions in Bose-Einstein Condensates via Optical Angular Momentum Beams*, Phys. Rev. Lett. **95**, 173601 (2005).
[37] B. P. Anderson, P. C. Haljan, C. A. Regal, D. L. Feder, L. A. Collins, C. W. Clark, and E. A. Cornell, *Watching Dark Solitons Decay into Vortex Rings in a Bose-Einstein Condensate*, Phys. Rev. Lett. **86**, 2926 (2001).
[38] N. S. Ginsberg, J. Brand, and L. V. Hau, *Observation of Hybrid Soliton Vortex-Ring Structures in Bose-Einstein Condensates*, Phys. Rev. Lett. **94**, 040403 (2005).
[39] S. Donadello, S. Serafini, M. Tylutki, L. P. Pitaevskii, F. Dalfovo, G. Lamporesi, and G. Ferrari, *Observation of Solitonic Vortices in Bose-Einstein Condensates*, Phys. Rev. Lett. **113**, 065302 (2014).
[40] S. Burger, K. Bongs, S. Dettmer, W. Ertmer, K. Sengstock, A. Sanpera, G. V. Shlyapnikov, and M. Lewenstein, *Dark Solitons in Bose-Einstein Condensates*, Phys. Rev. Lett. **83**, 5198 (1999).
[41] C.-A. Chen and C.-L. Hung, *Observation of Universal Quench Dynamics and Townes Soliton Formation from Modulational Instability in Two-Dimensional Bose Gases*, Phys. Rev. Lett. **125**, 250401 (2020).
[42] S.W. Seo, B. Ko, J. H. Kim, and Y. Shin, Observation of vortex-antivortex pairing in decaying 2D turbulence of a superfluid gas, Sci. Rep. **7**, 4587 (2017).
[43] P. C. Haljan, B. P. Anderson, I. Coddington, and E. A. Cornell, *Use of Surface-Wave Spectroscopy to Characterize Tilt Modes of a Vortex in a Bose-Einstein Condensate*, Phys. Rev. Lett. **86**, 2922 (2001).
[44] A. T. Powis, S. J. Sammut, and T. P. Simula, *Vortex Gyroscope Imaging of Planar Superfluids*, Phys. Rev. Lett. **113**, 165303 (2014).
[45] R. N. Bisset, S. Serafini, E. Iseni, M. Barbiero, T. Bienaimé, G. Lamporesi, G. Ferrari, and F. Dalfovo, *Observation of a Spinning Top in a Bose-Einstein Condensate*, Phys. Rev. A **96**, 053605 (2017).
[46] S. Serafini, L. Galantucci, E. Iseni, T. Bienaimé, R. N. Bisset, C. F. Barenghi, F. Dalfovo, G. Lamporesi, and G. Ferrari, *Vortex Reconnections and Rebounds in Trapped Atomic Bose-Einstein Condensates*, Phys. Rev. X **7**, 021031 (2017).
[47] M. Tylutki, S. Donadello, S. Serafini, L. P. Pitaevskii, F. Dalfovo, G. Lamporesi, and G. Ferrari, *Solitonic Vortices in Bose–Einstein Condensates*, Eur. Phys. J. Spec. Top. **224**, 577 (2015).
[48] F. Dalfovo, S. Giorgini, L. P. Pitaevskii, and S. Stringari, *Theory of Bose-Einstein Condensation in Trapped Gases*, Rev. Mod. Phys. **71**, 463 (1999).
[49] C. J. Pethick and H. Smith, Bose–Einstein Condensation in Dilute Gases (Cambridge University Press, Cambridge, England, 2008).
[50] A. V. Oppenheim and J. S. Lim, *The Importance of Phase in Signals*, Proceedings of the IEEE **69**, 529 (1981).

[51] S. Marchesini, Invited article: A unified evaluation of iterative projection algorithms for phase retrieval, Rev. Sci. Instrum. **78**, 011301 (2007).

[52] C.-H. Lu, C. Barsi, M. O. Williams, J. N. Kutz, and J. W. Fleischer, *Phase Retrieval Using Nonlinear Diversity*, Applied Optics **52**, D92 (2013).

[53] S. Asokan, P. A. Yasir, and J. S. Ivan, *Estimation of Dislocated Phases in Wavefronts through Intensity Measurements Using a Gerchberg–Saxton Type Algorithm*, Applied Optics **59**, 7225 (2020).

[54] J. W. Fleischer, G. Bartal, O. Cohen, O. Manela, M. Segev, J. Hudock, and D. N. Christodoulides, *Observation of Vortex-Ring ``Discrete'' Solitons in 2D Photonic Lattices*, Phys. Rev. Lett. **92**, 123904 (2004).

[55] N. P. Proukakis and B. Jackson, *Finite-Temperature Models of Bose–Einstein Condensation*, J. Phys. B: At. Mol. Opt. Phys. **41**, 203002 (2008).

[56] H. Buljan, M. Segev, and A. Vardi, *Incoherent Matter-Wave Solitons and Pairing Instability in an Attractively Interacting Bose-Einstein Condensate*, Phys. Rev. Lett. **95**, 180401 (2005).

[57] R. Ziv, H. Tamura, Y. Sagi, C.-L. Hung, and M. Segev, Experimental phase retrieval of matter waves, in 2023 Conference on Lasers and Electro-Optics (CLEO) (IEEE, San Jose, CA, 2023), pp. 1–2.

[58] A. Luski, Y. Segev, R. David, O. Bitton, H. Nadler, A. R. Barnea, A. Gorlach, O. Cheshnovsky, I. Kaminer, and E. Narevicius, *Vortex Beams of Atoms and Molecules*, Science **373**, 1105 (2021).

**Supplementary material A – Optimal Propogation Time Under Partial Trap**

We wish to solve the following problem and find the optimal $T_1$ parameter maximizing the difference between the magnitude of the $\pm 1$ vortex states after propagation in our measurement scheme:

$$\max_{T_1} \left\| |\psi_1(x,y,T_1+T_2)|^2 - |\psi_{-1}(x,y,T_1+T_2)|^2 \right\|$$

We consider three Hamiltonians relating to the three stages of the augmented TOF measurement in the linear regime. $H_0$ the Hamiltonian during complete trapping of the condensate, $H_1$ after opening one axis of the trap and $H_2$ the Hamiltonian of free propagation:

$$H_0 = \frac{P^2}{2m} + V(x,y) \quad H_1 = \frac{P^2}{2m} + V(x) \quad H_2 = \frac{P^2}{2m}$$

Let us take a symmetric separable potential. Since the potential is separable the solution is separable as well [1] and the eigen basis for the x and y dimensions is the same due to the symmetry of the Hamiltonian. A vortex mode is then given by:

$$\psi_{\pm 1} = u_0(x)u_1(y) \pm i u_1(x)u_0(y)$$

where $u_i(\cdot)$ is the i-th eigenfunction of the one-dimensional Hamiltonian, $(\cdot)$ indicates the dependence on the dimension coordinate ($x/y$ in our case). If we lift the potential barrier in one direction, the eigenfunction basis in that direction changes to that of a plane wave. We have:

$$u_i(\cdot) = \int_{-\infty}^{\infty} \langle k, u_i \rangle |k\rangle dk = \int_{-\infty}^{\infty} e^{ik(\cdot)} dk \int_{-\infty}^{\infty} u_i(\cdot) e^{-ik(\cdot)} d(\cdot)$$

which is the Fourier basis. Propagation in time is given by (for brevity $\hbar$ is omitted and we implicitly divide and rescale our Hamiltonians by $\hbar$):

$$\psi(x,y,t_0+T) = e^{-iH_i T} \psi(x,y,t_0)$$

If the wavefunction is an eigenfunction with eigenvalue (energy) we have:

$$\psi(x,y,t_0+T) = e^{-iH_i T} \psi(x,y,t_0) = e^{-iE_i T} \psi(x,y,t_0)$$

Therefore:

$$e^{-iH_1 T_1} \psi_{\pm 1} = e^{-iH_1 T_1} \left( u_0(x)u_1(y) \pm i u_1(x)u_0(y) \right)$$

$$= \left( e^{-iE_0 T_1} u_0(x) \int_{-\infty}^{\infty} e^{-i\frac{k^2}{2m}T_1} \langle k, u_1 \rangle |k\rangle dk \pm i e^{-iE_1 T_1} u_1(x) \int_{-\infty}^{\infty} e^{-i\frac{k^2}{2m}T_1} \langle k, u_0 \rangle |k\rangle dk \right)$$

$$= \left( e^{-iE_0 T_1} u_0(x) \int_{-\infty}^{\infty} u_1^{\mathcal{F}}(k) e^{-i\frac{k^2}{2m}T_1} e^{iky} dk \pm i e^{-iE_1 T_1} u_1(x) \int_{-\infty}^{\infty} u_0^{\mathcal{F}}(k) e^{-i\frac{k^2}{2m}T_1} e^{iky} dk \right)$$

Here, $u_i^{\mathcal{F}}(k)$ represents the Fourier coefficient of the i-th eigenfunction.

After propagation, a time $T_1$ under potential in one direction, we remove the second potential and get:

$$e^{-iH_2T_2}\psi_{\pm 1} = \begin{pmatrix} e^{-iE_0T_1}\int_{-\infty}^{\infty}u_0^{\mathcal{F}}(k)e^{-i\frac{k^2}{2m}T_2}e^{ikx}dk \cdot \int_{-\infty}^{\infty}u_1^{\mathcal{F}}(k)e^{-i\frac{k^2}{2m}(T_1+T_2)}e^{iky}dk \\ \pm ie^{-iE_1T_1}\int_{-\infty}^{\infty}u_1^{\mathcal{F}}(k)e^{-i\frac{k^2}{2m}T_2}e^{ikx}dk\int_{-\infty}^{\infty}u_0^{\mathcal{F}}(k)e^{-i\frac{k^2}{2m}(T_1+T_2)}e^{iky}dk \end{pmatrix}$$

Substituting in for the norm difference we wish to evaluate:

$$\left\||\psi_1|^2 - |\psi_{-1}|^2\right\|$$

Now,

$$\left\||\psi_1|^2 - |\psi_{-1}|^2\right\| = \left\| \begin{vmatrix} e^{-iE_0T_1}\int_{-\infty}^{\infty}u_0^{\mathcal{F}}(k)e^{-i\frac{k^2}{2m}T_2}e^{ikx}dk \cdot \int_{-\infty}^{\infty}u_1^{\mathcal{F}}(k)e^{-i\frac{k^2}{2m}(T_1+T_2)}e^{iky}dk \\ +ie^{-iE_1T_1}\int_{-\infty}^{\infty}u_1^{\mathcal{F}}(k)e^{-i\frac{k^2}{2m}T_2}e^{ikx}dk\int_{-\infty}^{\infty}u_0^{\mathcal{F}}(k)e^{-i\frac{k^2}{2m}(T_1+T_2)}e^{iky}dk \end{vmatrix}^2 \right. \\ \left. - \begin{vmatrix} e^{-iE_0T_1}\int_{-\infty}^{\infty}u_0^{\mathcal{F}}(k)e^{-i\frac{k^2}{2m}T_2}e^{ikx}dk \cdot \int_{-\infty}^{\infty}u_1^{\mathcal{F}}(k)e^{-i\frac{k^2}{2m}(T_1+T_2)}e^{iky}dk \\ -ie^{-iE_1T_1}\int_{-\infty}^{\infty}u_1^{\mathcal{F}}(k)e^{-i\frac{k^2}{2m}T_2}e^{ikx}dk\int_{-\infty}^{\infty}u_0^{\mathcal{F}}(k)e^{-i\frac{k^2}{2m}(T_1+T_2)}e^{iky}dk \end{vmatrix}^2 \right\|$$

In order to evaluate the integral, we approximate its value using the stationary phase approximation [2]:

$$\int_{-\infty}^{\infty}u_i^{\mathcal{F}}(k)e^{-i\frac{k^2}{2m}T_2}e^{ik(\cdot)}dk \approx u_i^{\mathcal{F}}\left(\frac{m(\cdot)}{T_2}\right)e^{i\frac{m(\cdot)^2}{2T_2}}\int_{-\infty}^{\infty}e^{-i\frac{T_2}{2m}\left(k-\frac{m(\cdot)}{T_2}\right)^2}dk = u_i^{\mathcal{F}}\left(\frac{m(\cdot)}{T_2}\right)e^{i\frac{m(\cdot)^2}{2T_2}}\sqrt{-\frac{2\pi mi}{T_2}}$$

Substituting this into the previous equation:

$$\left\||\psi_1|^2 - |\psi_{-1}|^2\right\| = \left\| \begin{vmatrix} e^{-iE_0T_1}u_0^{\mathcal{F}}\left(\frac{mx}{T_2}\right)e^{i\frac{mx^2}{2T_2}}\sqrt{-\frac{2\pi mi}{T_2}} \cdot u_1^{\mathcal{F}}\left(\frac{my}{(T_1+T_2)}\right)e^{i\frac{my^2}{2(T_1+T_2)}}\sqrt{-\frac{2\pi mi}{(T_1+T_2)}} \\ +ie^{-iE_1T_1}u_1^{\mathcal{F}}\left(\frac{mx}{T_2}\right)e^{i\frac{mx^2}{2T_2}}\sqrt{-\frac{2\pi mi}{T_2}}u_0^{\mathcal{F}}\left(\frac{my}{(T_1+T_2)}\right)e^{i\frac{my^2}{2(T_1+T_2)}}\sqrt{-\frac{2\pi mi}{(T_1+T_2)}} \end{vmatrix}^2 \right. \\ \left. - \begin{vmatrix} e^{-iE_0T_1}u_0^{\mathcal{F}}\left(\frac{mx}{T_2}\right)e^{i\frac{mx^2}{2T_2}}\sqrt{-\frac{2\pi mi}{T_2}} \cdot u_1^{\mathcal{F}}\left(\frac{my}{(T_1+T_2)}\right)e^{i\frac{my^2}{2(T_1+T_2)}}\sqrt{-\frac{2\pi mi}{(T_1+T_2)}} \\ -ie^{-iE_1T_1}u_1^{\mathcal{F}}\left(\frac{mx}{T_2}\right)e^{i\frac{mx^2}{2T_2}}\sqrt{-\frac{2\pi mi}{T_2}}u_0^{\mathcal{F}}\left(\frac{my}{(T_1+T_2)}\right)e^{i\frac{my^2}{2(T_1+T_2)}}\sqrt{-\frac{2\pi mi}{(T_1+T_2)}} \end{vmatrix}^2 \right\|$$

Removing the common phase term and taking the common factor outside of the norm operation yields:

$$= \frac{2\pi m}{\sqrt{T_2(T_1+T_2)}} \left\| \begin{array}{c} \left| -e^{-iE_0T_1} u_0^{\mathcal{F}}\left(\frac{mx}{T_2}\right) \cdot u_1^{\mathcal{F}}\left(\frac{my}{(T_1+T_2)}\right) i + e^{-iE_1T_1} u_1^{\mathcal{F}}\left(\frac{mx}{T_2}\right) u_0^{\mathcal{F}}\left(\frac{my}{(T_1+T_2)}\right) \right|^2 \\ - \left| -e^{-iE_0T_1} u_0^{\mathcal{F}}\left(\frac{mx}{T_2}\right) \cdot u_1^{\mathcal{F}}\left(\frac{my}{(T_1+T_2)}\right) i + e^{-iE_1T_1} u_1^{\mathcal{F}}\left(\frac{mx}{T_2}\right) u_0^{\mathcal{F}}\left(\frac{my}{(T_1+T_2)}\right) \right|^2 \end{array} \right\|$$

We note that the eigen wavefunctions can be taken to be real, and by the properties of bounded eigen wavefunctions in symmetric potentials, the ground state is even and the first excited is odd. Therefore, by the Fourier properties, in the Fourier plane $u_0^{\mathcal{F}}(\cdot)$ is a real and even function and $u_1^{\mathcal{F}}(\cdot)$ is an imaginary and even function. Therefore:

$$\frac{2\pi m}{\sqrt{T_2(T_1+T_2)}} \left\| \begin{array}{c} \left| u_0^{\mathcal{F}}\left(\frac{mx}{T_2}\right) \cdot u_1^{\mathcal{F}}\left(\frac{my}{(T_1+T_2)}\right) + u_1^{\mathcal{F}}\left(\frac{mx}{T_2}\right) u_0^{\mathcal{F}}\left(\frac{my}{(T_1+T_2)}\right) (i\cos(\Delta E T_1) + \sin(\Delta E T_1)) \right|^2 \\ - \left| -u_0^{\mathcal{F}}\left(\frac{mx}{T_2}\right) \cdot u_1^{\mathcal{F}}\left(\frac{my}{(T_1+T_2)}\right) + u_1^{\mathcal{F}}\left(\frac{mx}{T_2}\right) u_0^{\mathcal{F}}\left(\frac{my}{(T_1+T_2)}\right) (i\cos(\Delta E T_1) + \sin(\Delta E T_1)) \right|^2 \end{array} \right\|$$

By the absolute value operation:

$$= \frac{2\pi m}{\sqrt{T_2(T_1+T_2)}} \left\| \begin{array}{c} \left( \left( u_0^{\mathcal{F}}\left(\frac{mx}{T_2}\right) \cdot u_1^{\mathcal{F}}\left(\frac{my}{(T_1+T_2)}\right) + u_1^{\mathcal{F}}\left(\frac{mx}{T_2}\right) u_0^{\mathcal{F}}\left(\frac{my}{(T_1+T_2)}\right) \sin(\Delta E T_1) \right)^2 \\ + \left( u_1^{\mathcal{F}}\left(\frac{mx}{T_2}\right) u_0^{\mathcal{F}}\left(\frac{my}{(T_1+T_2)}\right) \cos(\Delta E T_1) \right)^2 \\ - \left( \left( u_0^{\mathcal{F}}\left(\frac{mx}{T_2}\right) \cdot u_1^{\mathcal{F}}\left(\frac{my}{(T_1+T_2)}\right) - u_1^{\mathcal{F}}\left(\frac{mx}{T_2}\right) u_0^{\mathcal{F}}\left(\frac{my}{(T_1+T_2)}\right) \sin(\Delta E T_1) \right)^2 \right. \\ \left. + \left( u_1^{\mathcal{F}}\left(\frac{mx}{T_2}\right) u_0^{\mathcal{F}}\left(\frac{my}{(T_1+T_2)}\right) \cos(\Delta E T_1) \right)^2 \right) \end{array} \right\|$$

Simplifying, we obtain:

$$= \frac{2\pi m}{\sqrt{T_2(T_1+T_2)}} \left\| \begin{array}{c} \left( u_0^{\mathcal{F}}\left(\frac{mx}{T_2}\right) \cdot u_1^{\mathcal{F}}\left(\frac{my}{(T_1+T_2)}\right) + u_1^{\mathcal{F}}\left(\frac{mx}{T_2}\right) u_0^{\mathcal{F}}\left(\frac{my}{(T_1+T_2)}\right) \sin(\Delta E T_1) \right)^2 \\ - \left( u_0^{\mathcal{F}}\left(\frac{mx}{T_2}\right) \cdot u_1^{\mathcal{F}}\left(\frac{my}{(T_1+T_2)}\right) - u_1^{\mathcal{F}}\left(\frac{mx}{T_2}\right) u_0^{\mathcal{F}}\left(\frac{my}{(T_1+T_2)}\right) \sin(\Delta E T_1) \right)^2 \end{array} \right\|$$

$$= \frac{2\pi m}{\sqrt{T_2(T_1+T_2)}} \left\| 4 u_0^{\mathcal{F}}\left(\frac{mx}{T_2}\right) \cdot u_1^{\mathcal{F}}\left(\frac{my}{(T_1+T_2)}\right) \cdot u_1^{\mathcal{F}}\left(\frac{mx}{T_2}\right) u_0^{\mathcal{F}}\left(\frac{my}{(T_1+T_2)}\right) \sin(\Delta E T_1) \right\|$$

Finally, we find:

$$\left\| |\psi_1|^2 - |\psi_{-1}|^2 \right\| = \frac{2\pi m}{\sqrt{T_2(T_1+T_2)}} \left\| 4u_0^{\mathcal{F}}\left(\frac{mx}{T_2}\right) \cdot u_1^{\mathcal{F}}\left(\frac{my}{(T_1+T_2)}\right) \cdot u_1^{\mathcal{F}}\left(\frac{mx}{T_2}\right) u_0^{\mathcal{F}}\left(\frac{my}{(T_1+T_2)}\right) \sin(\Delta E T_1) \right\|$$

We see that the magnitude difference is proportional to $\frac{\sin(\Delta E T_1)}{\sqrt{T_2(T_1+T_2)}}$ directly, with indirect dependence on $T_1$ through the scaling of $u_1^{\mathcal{F}}, u_0^{\mathcal{F}}$ in the norm operation. As the scaling is governed by $T_2$, which we assume to be larger than $T_1$ in our setting, the overlap between the eigenfunctions in the norm operation has weak dependence on $T_1$. We conclude that:

$$\left\| |\psi_1|^2 - |\psi_{-1}|^2 \right\| \propto \left| \frac{\sin(\Delta E T_1)}{\sqrt{T_2(T_1+T_2)}} \right|$$

[1]   L. P. Eisenhart, Enumeration of potentials for which one particle Schrödinger equations are separable, Phys. Rev. A 74, 87 (1948).
[2]   R. Wong, Asymptotics Approximation of Integrals, SIAM Classics in Applied Mathematics (SIAM, Philadelphia, 2001).

**Supplemanty material B – Simulation detalis**

**Vortex and Latice details**

In this section, we describe the simulation method and parameters used in the main text. We solve a dimensionless GPE:

$$\left( -\frac{\hbar^2}{2m}\nabla^2 + V(\mathbf{r}) + U_0 |\psi(\mathbf{r},t)|^2 \right) \psi(\mathbf{r},t) = i\hbar \partial_t \psi(\mathbf{r},t)$$

where $\hbar, m$ are taken to be unity. The potential is taken as a potential well with depth $10 [arb.units]$ and radius 1. The nonlinear coefficient is chosen in the range $U_0 \in [0,100]$. As the wavefunction is normalized to unity, it should be understood that this factor implicitly contains the total particle number inside the condensate, i.e., $U_0 = \frac{4\pi\hbar^2 a_s}{m} N_0$, where $N_0$ is the total particle number and $a_s$ is the scattering length. We note that in Fig 2 and Fig 3, $U_0$ is taken as 0 for linear propagation, 100 for weak nonlinearity, and 10000 for strong nonlinearity. These values are chosen such that $V(\mathbf{r}) \gg U_0 |\psi(\mathbf{r},t)|^2$ and $U_0 |\psi(\mathbf{r},t)|^2 \gg V(\mathbf{r})$, for the last two cases, respectively.

The spatial dimensions are discretized into a $1024 \times 1024$ grid with a resolution of $dx = dy = 0.0756 [arb.units]$. For propagation, we use the BPM or split-step method with a temporal step of $dt = 10^{-3} [arb.units]$. The number of steps for complete propagation in the TOF and augmented TOF is taken to be 3000. In the augmented TOF, the number of steps taken under the partial propagation is set to be 558 (corresponding to $\Delta E T_1/\hbar = \pi/2$, see section 3.3), while the rest of the propagation steps are done under free propagation with no potential. We note that, for the noise simulations the propagation time is taken to be slightly shorter in order to ease computation time over repeated runs, a total of 1500 steps with 558 being under the partial potential and the rest in free propagation.

The lattice is constructed by duplicating the well potential into a 3x3 lattice with separation of $\Delta x = \Delta y \approx 2.4 [arb.units]$. The vortex lattice wavefunction is constructed in a similar way, placing a single vortex at the center of each well potential with a random flow direction and relative phase to the other sites.

**Random phase**

The simulation details for the random phase are similar for the above case with the following parameter changes: The trap was taken as a harmonic trap, $V(x,y) = \frac{1}{2}(x^2 + y^2)$, the spatial discretization $dx = dy = 0.0938 [arb.units]$, $U_0 \cong 8000 [arb.units]$ (strong nonlinearity) and $dt = 2 \times 10^{-5} [arb.units]$. Complete propagation in the augmented TOF is taken to be 513, with 200 steps under partial propagation. The random phase pattern was generated by generating an initial Gaussian noise image, filtering it in the Fourier plane to create smooth phase pattern and scaling its values to be in the $[0, 2\pi]$ range.